\documentclass[conference]{IEEEtran}

\IEEEoverridecommandlockouts
\usepackage{verbatim}
\usepackage{bm}
\usepackage{longtable}
\DeclareMathAlphabet{\mathpzc}{OT1}{pzc}{m}{it}
\usepackage{amsmath}
\usepackage{amsfonts}
\usepackage{amssymb}
\usepackage{multirow}
\usepackage{amsmath,amssymb,amsfonts}
\usepackage{fancyhdr}
\usepackage[ruled,vlined,linesnumbered]{algorithm2e}
\usepackage[
top    = 0.7in,
bottom = 0.99in,
left   = 0.65in,
right  = 0.59in]{geometry}
\usepackage{graphicx}
\usepackage{textcomp}
\usepackage{xcolor}
\usepackage{subcaption}
\usepackage{algpseudocode}
\usepackage{enumitem}
\usepackage{float}
\restylefloat{figure}
\usepackage{graphicx}
\usepackage[normalem]{ulem}
\useunder{\uline}{\ul}{}

\begin{document}

\title{Impact of Conflicting Transactions in Blockchain: Detecting and Mitigating Potential Attacks}

\author{\IEEEauthorblockN{Faisal Haque Bappy$^{1}$, Tariqul Islam$^{2}$, Kamrul Hasan$^{3}$, Joon S. Park$^{4}$, and Carlos Caicedo$^{5}$}
\IEEEauthorblockA{
$^{1, 2, 4, 5}$ Syracuse University, Syracuse, NY, USA\\
$ ^{3}$ Tennessee State University, Nashville, TN, USA\\
Email: \{fbappy@syr, mtislam@syr, mhasan1@tnstate, jspark@syr, ccaicedo@syr\}.edu} 
}

\maketitle

\thispagestyle{fancy}
 \lhead{This work has been accepted at the 2024 IEEE Global Communications Conference (Globecom 2024)}
\cfoot{}

\begin{abstract}
Conflicting transactions within blockchain networks not only pose performance challenges but also introduce security vulnerabilities, potentially facilitating malicious attacks. In this paper, we explore the impact of conflicting transactions on blockchain attack vectors. Through modeling and simulation, we delve into the dynamics of four pivotal attacks - block withholding, double spending, balance, and distributed denial of service (DDoS), all orchestrated using conflicting transactions. Our analysis not only focuses on the mechanisms through which these attacks exploit transaction conflicts but also underscores their potential impact on the integrity and reliability of blockchain networks. Additionally, we propose a set of countermeasures for mitigating these attacks. Through implementation and evaluation, we show their effectiveness in lowering attack rates and enhancing overall network performance seamlessly, without introducing additional overhead. Our findings emphasize the critical importance of actively managing conflicting transactions to reinforce blockchain security and performance.

\end{abstract}


\section{Introduction}
Conflicting transactions within blockchain networks are a prevalent phenomenon, occurring in both permissioned and permissionless environments due to the complex interactions among multiple wallets or transactions \cite{amiri2019parblockchain}. While often considered as bottlenecks for performance, these conflicts harbor deeper security implications, potentially facilitating various malicious activities. In a blockchain network, conflicting transactions occur when multiple transactions attempt to modify the same asset simultaneously. These conflicts arise from factors such as network latency, transaction competition, and deliberate manipulation. In permissioned blockchains, conflicts may arise due to concurrent transaction execution by different participants \cite{nasirifard2019fabriccrdt}, while in permissionless blockchains, they result from the race condition inherent in the consensus mechanism.

Despite their understated significance, conflicting transactions have profound implications for blockchain security. They serve as a breeding ground for attacks such as double spending, block withholding, balance, and distributed denial of service (DDoS), undermining the integrity and trustworthiness of blockchain systems. The double spending attack involves attempting to spend the same cryptocurrency tokens twice by creating conflicting transactions targeting different recipients \cite{zhang2019double}. Attackers can exploit the temporal gap between transaction validation and confirmation to deceive the network into accepting both transactions, defrauding legitimate recipients. Block withholding attacks disrupt block validation and propagation, aiming to compromise the reliability of the consensus mechanism \cite{lee2019countering, feng2021security}. Attackers can strategically withhold validated blocks containing conflicting transactions, inducing inconsistencies and potential forks in the network. Conflicting transactions can also manipulate account balances within blockchain networks \cite{feng2021security}, deceiving the network into updating balances inconsistently and causing financial losses for legitimate users. Despite extensive research on blockchain attacks \cite{gorenflo2020xox, gorenflo2020fastfabric}, the exploitation of conflicting transactions remains largely unexplored. While existing studies focus on attack vectors like selfish mining and smart contract vulnerabilities \cite{saad2020exploring}, the potential of conflicting transactions as a tool for malicious actors has been overlooked. These conflicting transactions, when deployed strategically, can disrupt consensus mechanisms, compromise transaction integrity, and undermine network security.

To address this issue, we modeled and simulated four major types of attacks in this paper to uncover the vulnerabilities linked with conflicting transactions and assess their potential impact on blockchain security. Furthermore, we proposed countermeasures to mitigate the attacks and evaluated their efficacy in a controlled testbed environment. The following are the main contributions of this paper.

\begin{itemize}
    \item We presented the impact of conflicting transactions within the blockchain network, underscoring their pivotal role in extending potential attack surfaces.
    \item We modeled and simulated block withholding, double spending, balance, and DDoS attacks utilizing conflicting transactions. 
    \item We proposed countermeasures against these attacks which effectively decrease attack rates and improve the overall network performance.
\end{itemize}

The remainder of the paper is structured as follows: Section \ref{background} covers preliminary terms and concepts. In section \ref{attack-model}, we present attack models and simulation results. In Section \ref{counter}, we propose countermeasures for the attacks, which are then implemented and evaluated in Section \ref{evaluation}. Section \ref{related} presents existing work, followed by Section \ref{conclude}, which concludes the paper.

\section{Preliminaries}
\label{background}
\subsection{Permissioned and Permissionless Blockchain}
Permissioned blockchains, common in enterprise settings, restrict participation to authorized entities, offering controlled access, privacy features, and efficient governance mechanisms. Examples include Hyperledger Fabric \cite{HyperledgerFabric} and Corda \cite{corda}. Conversely, permissionless blockchains like Bitcoin and Ethereum allow anyone to join and participate, emphasizing decentralization, transparency, and censorship resistance. These networks rely on incentives, immutability, and resilience to foster trust and integrity. The choice between permissioned and permissionless blockchains depends on factors such as trust requirements, scalability needs, and desired levels of decentralization, each tailored to specific use cases and industries within the blockchain ecosystem.

\subsection{Conflicting Transactions in Blockchain}
\begin{figure}[tbh]
\centering
\includegraphics[width=\columnwidth]{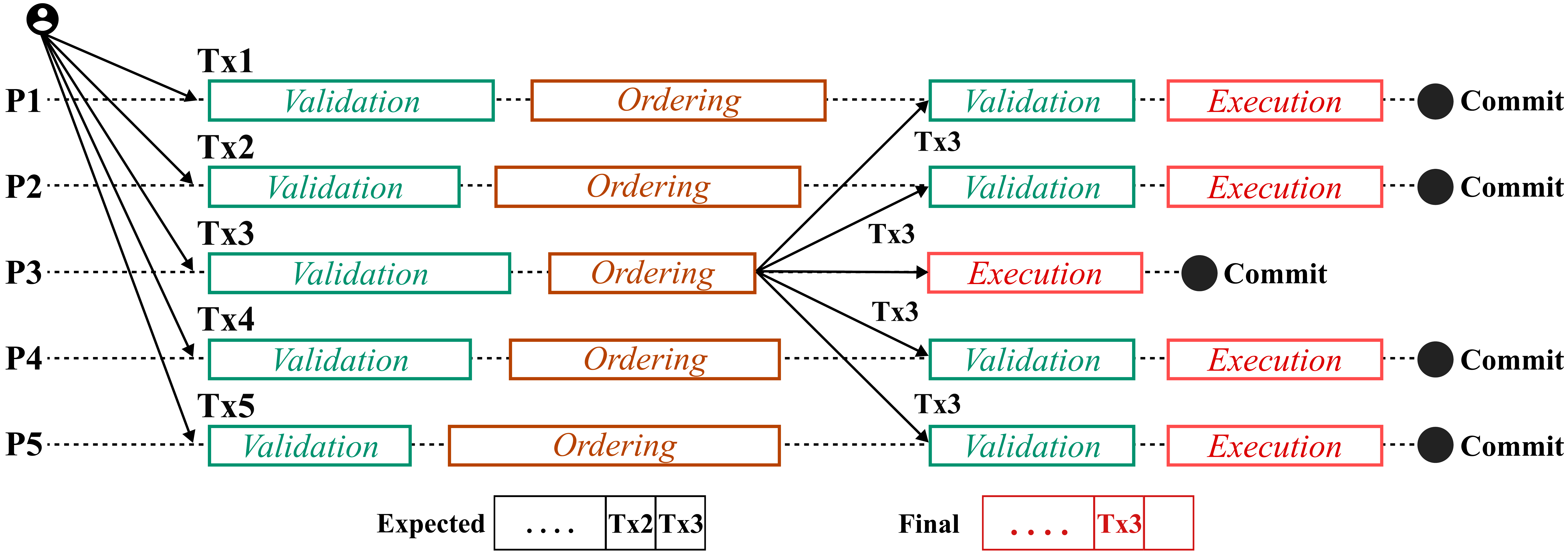}
        \caption{Conflicting Transactions in Blockchain} \label{fig:contention}
\end{figure}

In the context of blockchain, conflicting transaction refers to the occurrence of multiple simultaneous transactions attempting conflicting operations on a single record, typically a previous block or wallet. When a new transaction arrives from the client, it undergoes initial validation before peers begin the ordering process. Once ordered, transactions are executed across all peers, with the new block committed to the main chain. However, if multiple transactions arrive simultaneously, as illustrated in Figure \ref{fig:contention}, there is a risk of inconsistency \cite{amiri2019parblockchain}. A client submitted five transaction requests: Tx1, Tx2, Tx3, Tx4, and Tx5. Notably, Tx3 relies on Tx2 for consistency, necessitating Tx2 to be executed before Tx3. However, with simultaneous arrival, peers individually validate and order transactions. If peer 3 ($P_3$) orders Tx3 before others, it broadcasts the transaction, causing execution to begin prematurely on other peers. This results in an inconsistency as Tx3 should follow Tx2. Consequently, this may lead to an execution error or render the chain invalid. In this scenario, both Tx2 and Tx3 face conflicts due to their dependencies. In large-scale networks, such conflicting transactions are significant, potentially causing delays and impeding transaction processing.

\section{Attack Models}
\label{attack-model}

In this section, we investigate four types of attacks that can be performed by exploiting conflicting transactions. We presented the attack model for block withholding, double spending, balance, and DDoS attacks. Also, we simulated all of these attacks in a Hyperledger Fabric blockchain. For demonstrating the workflow of the attacks, we have used a minimal setup for fabric. Table \ref{tab:sim-env} shows the overview of our simulation environment. For generating conflicting transactions, we used the SmallBank dataset \cite{SmallBank} and synthetically created thousands of transactions that have read-and-write conflicts with several wallets (user accounts).

\begin{table}[tbh]
\centering
\caption{Overview of the Simulation Environment}
\label{tab:sim-env}
\begin{tabular}{ll}
\hline
\textbf{Configuration}                     & \textbf{Value/Description} \\ \hline
Platform                                   & Hyperledger Fabric         \\ \hline
Number of Nodes                            & 4, 8, 16, 32               \\ \hline
Consensus Algorithm                        & Raft                       \\ \hline
Smart Contract Language                    & GoLang                     \\ \hline
Transaction Type                           & Asset Transfer             \\ \hline
Asset Type                                 & Token                      \\ \hline
Number of Generated Conflicting Transactions         & 100                        \\ \hline
Time Span between Conflicting Transactions & 10 seconds                 \\ \hline
Network Latency Control                    & Virtual Network (VNet)     \\ \hline
Network Speed (Simulated)                  & 100 Mbps                   \\ \hline
Node CPU                                   & Single Core (2 GHz)        \\ \hline
Node RAM                                   & 2 GB                       \\ \hline
\end{tabular}
\end{table}

\subsection{Block Withholding Attack}
A block withholding attack involves withholding blocks containing valid transactions from the rest of the network while processing other blocks.

\textbf{Permissionless Model.}
In permissionless blockchains, an attacker can disrupt mining pools by holding blocks instead of publishing them, causing a significant loss in incentives. This attacker can act as both a user and a miner. The attacker deliberately submits numerous conflicting transactions to distract other miners from processing them \cite{feng2021security}, while retaining control over valid transactions.

\textbf{Permissioned Model.} 
In permissioned blockchains, attackers enter as regular users with the aim of disrupting transaction ordering. While block withholding attacks can be executed in various ways \cite{tosh2017security, feng2021security, qin2020optimal, saad2020exploring}, the main step for performing the attack is to determine which blocks to hold and trick the rest of the network into processing other blocks in the meantime. 

\begin{figure}[]
\centering
\includegraphics[width=0.8\columnwidth]{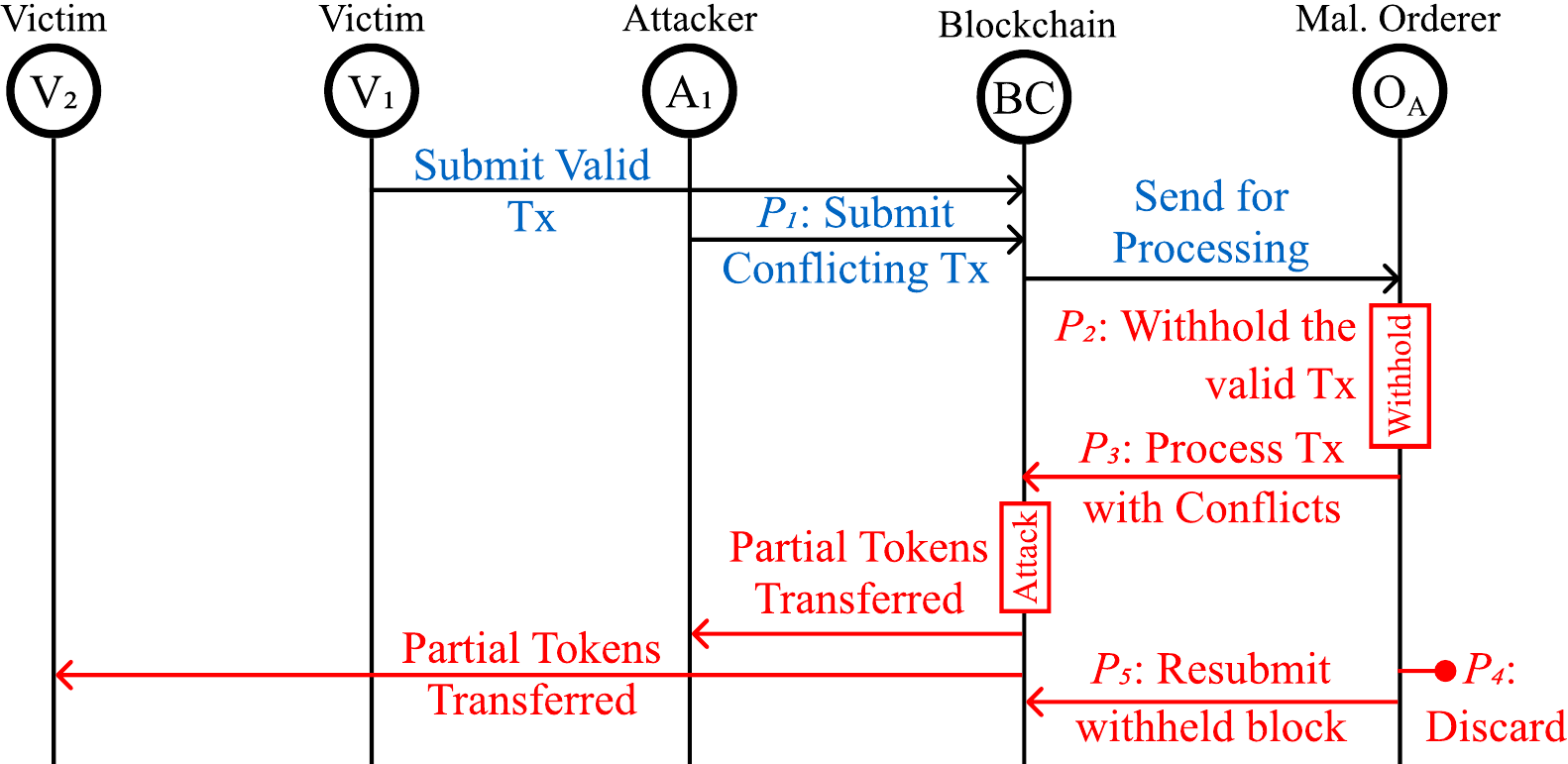}
        \caption{Sequence Diagram of Block Withholding Attack} \label{fig:bwh-pnet}
\end{figure}

\textbf{Workflow.}
For this attack, we have considered three wallets $A_1$, $V_1$, and $V_2$ with an initial balance of 1000 tokens each. The attacker owns wallet $A_1$ and the victims own the wallet $V_1$ and $V_2$. The attacker works as an orderer and the main goal here is to withhold a valid transaction from wallet $V_1$ to wallet $V_2$ and in the meantime manipulate the balances of other wallets by submitting and ordering conflicting transactions. Figure \ref{fig:bwh-pnet} shows the sequence diagram of the attack, where $P_1$, $P_2$, and $P_3$ are the pre-conditions for the attack while $P_4$ and $P_5$ are the post-conditions. $\bm{P_1}$: In this phase, the attacker focuses on a legitimate transaction, transferring 15 tokens from $V_1$ to $V_2$. Simultaneously, the attacker creates 100 conflicting transactions involving transfers from wallets $A_1$, $V_1$, and $V_2$. Intentionally, the attacker submits this entire batch of transactions simultaneously.
$\bm{P_2}$: In permissioned blockchains, the attacker, acting as an orderer, withholds the endorsed valid transaction once it meets requirements. Conversely, in permissionless blockchains, the attacker selects the targeted transaction from the mining pool and holds it after processing. As other transactions also got processed almost at the same time they will be queued for ordering and honest orderers will start ordering them.
$\bm{P_3}$: Here, the attacker initiates the ordering of conflicting transactions alongside the network, while retaining the targeted transaction. This action results in a substantial increase in transaction failures and artificially inflates the block height due to the influx of conflicting transactions. Now, with the successful withholding of the targeted block and execution of some conflicting transactions, the attacker has altered wallet balances. 
$\bm{P_4}$: If the attacker never submits the withheld block, the valid transaction will be marked as failed. This is likely to happen when the attacker gets the desired result from the attack, which is some balance increase to the wallet. Also, in permissionless blockchains, the whole mining pool will lose rewards in this case.
$\bm{P_5}$: If the attacker submits the withheld block, it can either get validated or invalidated based on the wallet balances and the order of blocks generated from conflicting transactions. 

\textbf{Outcome.}
After the block withholding attack, the final state (Table \ref{tab:result}) shows an artificial inflation of the block height to 105, encompassing 205 transactions. Due to conflicting transactions, wallets $A_1$ and $V_2$ see balance increases, while wallet $V_1$ experiences a decrease. This successful transfer of tokens to the attacker's wallets highlights the disruption caused by the attack and the potential for data manipulation within the blockchain network.


\subsection{Double Spending Attack}
A double spending attack happens when an attacker spends or uses a single asset multiple times. To perform this, the attacker submits a valid transaction first for transferring the asset to the victim. Then the attacker generates conflicting transactions which transfer the same asset to its other wallets \cite{duffield2014transaction}. The initial valid transaction may be discarded over time. If the attacker obtains the desired product or service from the victim during this window, the double spending attack is successful.

\textbf{Permissionless Model.} 
In permissionless blockchains, this attack scenario is highly dependent on the perfect timing and the ability to create delays for the initial transaction to be processed. The victim has to transfer the asset during that delay period.

\textbf{Permissioned Model.}
In permissioned blockchains, the participant numbers are relatively low compared to permissionless blockchains. So, the attacker has to submit more number of conflicting transactions to cause the delay.

\begin{figure}[tbh]
\centering
\includegraphics[width=0.85\columnwidth]{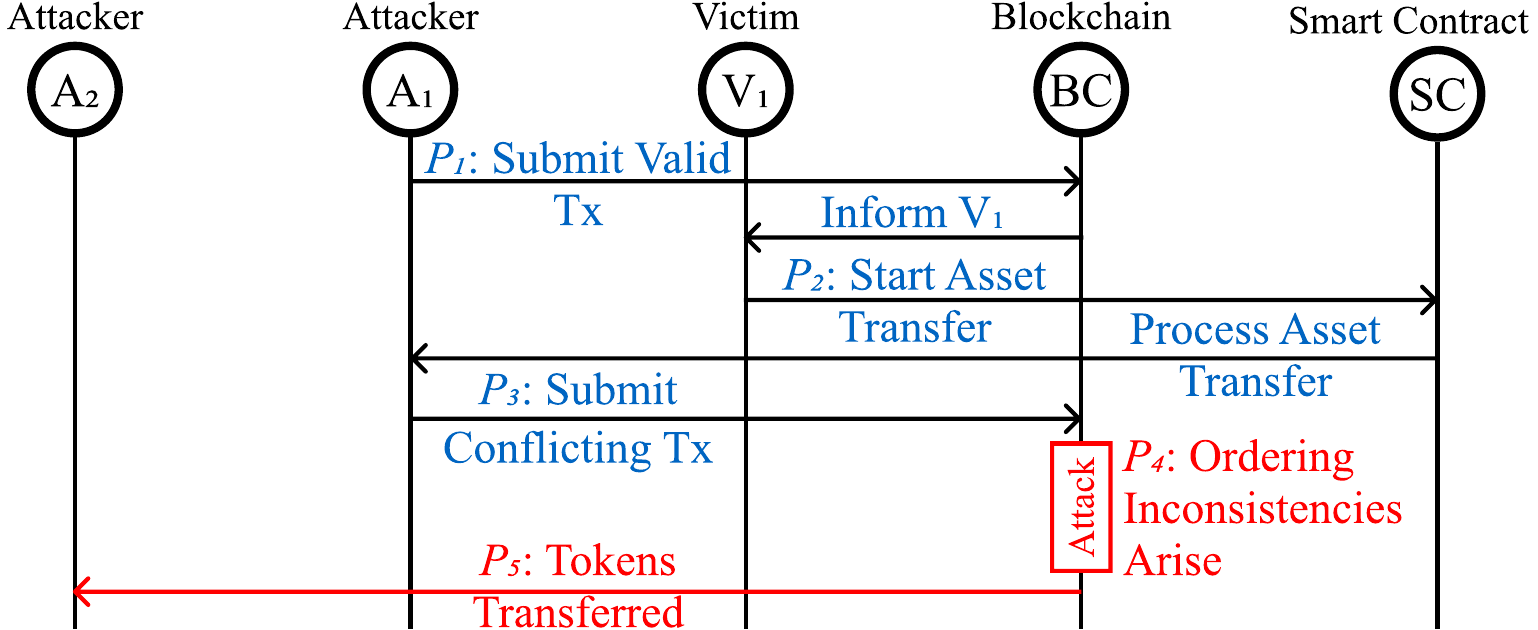}
        \caption{Sequence Diagram of Double Spending Attack} \label{fig:ds-pnet}
\end{figure}

\textbf{Workflow.}
In this attack scenario, we have three wallets: $A_1$, $V_1$, and $A_2$. The attacker possesses wallets $A_1$ and $A_2$, while the victim owns wallet $V_1$. We simulate a transaction where the attacker transfers 100 tokens to the victim's wallet, and the victim reciprocates through executing a smart contract or off-chain methods. However, the attacker aims to double-spend the 100 tokens, returning them to their other wallet by exploiting transaction ordering with conflicting transactions. Figure \ref{fig:ds-pnet} shows the sequence diagram of the attack, where $P_1$, $P_2$, $P_3$, and $P_4$ are the pre-conditions for the attack while $P_5$ is the post-condition for the attack.

$\bm{P_1}$: In this step, the attacker transfers 100 tokens from wallet $A_1$ to wallet $V_1$. After getting the required endorsements, the attacker assures the victim that the transaction is valid.
$\bm{P_2}$: After getting assurance for the initial transaction, the victim starts the asset transfer process by executing a smart contract that does not belong to the same channel.
$\bm{P_3}$: Now, the attacker gets confirmation about the asset transfer and tries to delay the ordering of the initial transaction by submitting 100 conflicting transactions which include wallets $A_1$, $A_2$, and $V_1$.
$\bm{P_4}$: Due to the conflicting transactions, the ordering process gets inconsistent. And as a result, some of the tokens get transferred from $A_1$ to $A_2$, instead of $V_1$. If the attacker gets confirmation of the successful transfer from $A_1$ to $A_2$, the attack is successful.
$\bm{P_5}$: Due to inconsistent ordering, one of the conflicting transactions ($A_1$ to $A_2$) was ordered and committed first. As a result, the initial valid transaction failed and the attacker regained the asset.

\textbf{Outcome.}
After the double spending attack, $A_1$'s balance was reduced to 0 tokens (Table \ref{tab:result}). $A_2$, on the other hand, received 100 tokens from $A_1$, so $A_2$'s balance after the attack increased to 100 units. But, victim $V_1$ did not receive any tokens from the transactions, while the asset transfer transaction was already executed. In this way, the attacker exploits the transaction ordering and double spends some tokens using conflicting transactions. 


\subsection{Balance Attack}
A balance attack happens when the attacker tries to create a fork for the main chain. The main goal of this attack is to break the main chain and perform replay attacks on the forked chain. 

\textbf{Permissionless Model.}
In permissionless blockchains, the attacker forms two subgroups with similar mining power and induces communication delays between them using conflicting transactions. This delay hampers one subgroup's block generation speed, causing the main chain to fork based on the longest-chain rule.

\textbf{Permissioned Model.}
In permissioned blockchains, the attacker focuses on channels rather than mining pools. Similar to permissionless models, the attacker joins both channels and introduces an unexpected communication delay between them. By exploiting this delay, the attacker can impede one channel from generating blocks at a faster pace.

\begin{figure}[tbh]
\centering
\includegraphics[width=0.85\columnwidth]{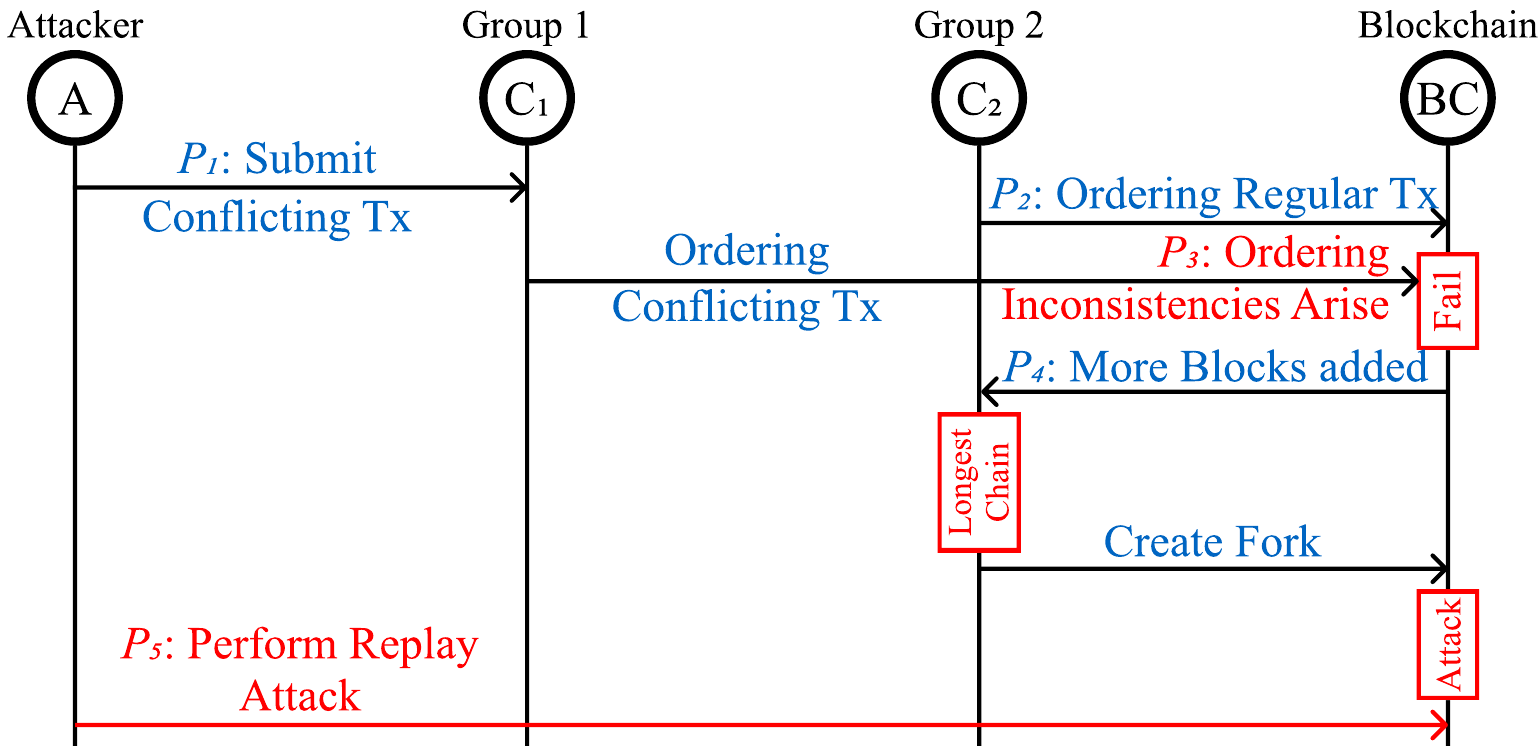}
        \caption{Sequence Diagram of Balance Attack} \label{fig:balance-pnet}
\end{figure}

\textbf{Workflow.}
For this attack, we have simulated a Hyperledger Fabric network with 2 channels $C_1$ and $C_2$. The attacker joins both $C_1$ and $C_2$, and then tries to delay the communication by submitting a lot of conflicting transactions in $C_1$. Figure \ref{fig:balance-pnet} shows the sequence diagram of the attack, where $P_1$, $P_2$, $P_3$, and $P_4$ are the pre-conditions for the attack while $P_5$ is the post-condition for the attack.
$\bm{P_1}$: In this step, the attacker submits 100 conflicting transactions in channel $C_1$ that replicates the valid transactions in $C_2$ but with conflicting wallets.  
$\bm{P_2}$: On the other hand, in $C_2$, the attacker continues ordering valid transactions. 
$\bm{P_3}$: Most of the transactions in $C_1$ fail and cause inconsistency in ordering. As a result, the valid transactions get delayed. 
$\bm{P_4}$: As $C_2$ continues to add valid blocks, it creates a chain that outweighs the one in $C_1$. The attacker successfully manipulates the network to approve a fork to the chain as $C_1$ is lagging behind $C_2$ due to conflicting transactions.
$\bm{P_5}$: After the attack is successful, the attacker double-spends or replays any transaction from the queue of the old chain in $C_1$. 

\textbf{Outcome.}
After the balance attack, the chain size of $C_2$ increases from 1000 to 1100, as indicated in Table \ref{tab:result}. Meanwhile, the chain size for $C_1$ remains at 1050. Additionally, the number of pending transactions in $C_1$ rises to 40 due to significant delays caused by conflicting transactions. Conversely, $C_2$ has no pending transactions post-attack. Therefore, if the attacker executes any pending transaction from $C_1$ in $C_2$, it can carry out a replay or double spending attack.


\begin{table}[]
\centering
\caption{Outcome of the Simulated Attacks}
\label{tab:result}
\begin{tabular}{lcc}
\hline

\multicolumn{3}{c}{\textbf{Block Withholding Attack}}                           \\ \hline
\textbf{Variable}               & \textbf{Initial State} & \textbf{Final State} \\ \hline
Block Height                    & 100                    & 105                  \\ \hline
Transaction Count               & 200                    & 205                  \\ \hline
Attacker's Wallet Balance $A_1$ & 1000                   & 1010                 \\ \hline
Victim's Wallet Balance $V_1$   & 1000                   & 985                  \\ \hline
Victim's Wallet Balance $V_2$   & 1000                   & 1005                 \\ \hline
                                                           \hline
\multicolumn{3}{c}{\textbf{Double Spending Attack}}                             \\ \hline
\textbf{Variable}               & \textbf{Initial State} & \textbf{Final State} \\ \hline
Attacker's Wallet Balance $A_1$ & 1000                   & 1010                 \\ \hline
Victim's Wallet Balance $V_1$   & 1000                   & 985                  \\ \hline
Victim's Wallet Balance $V_2$   & 1000                   & 1005                 \\ \hline
                                                          \hline
\multicolumn{3}{c}{\textbf{Balance Attack}}                                     \\ \hline
\textbf{Variable}               & \textbf{Initial State} & \textbf{Final State} \\ \hline
Chain Size in $C_1$             & 1000                   & 1050                 \\ \hline
Chain Size in $C_2$             & 1000                   & 1100                 \\ \hline
Pending Tx in $C_1$             & 10                     & 40                   \\ \hline
Pending Tx in $C_2$             & 10                     & 0                    \\ \hline
\end{tabular}
\end{table}

\subsection{DDoS Attack}
In DDoS attacks, attackers can target specific peers or the whole network. Every blockchain has a limit for the mempool (i.e., a contraction of memory and pool) and ordering queue \cite{saad2019mempool}. Being a distributed network it already has some latency for communicating between all peers. So, an attacker can make use of the conflicting transactions here. 

\begin{figure}[b]
\centering
\includegraphics[width=0.9\columnwidth]{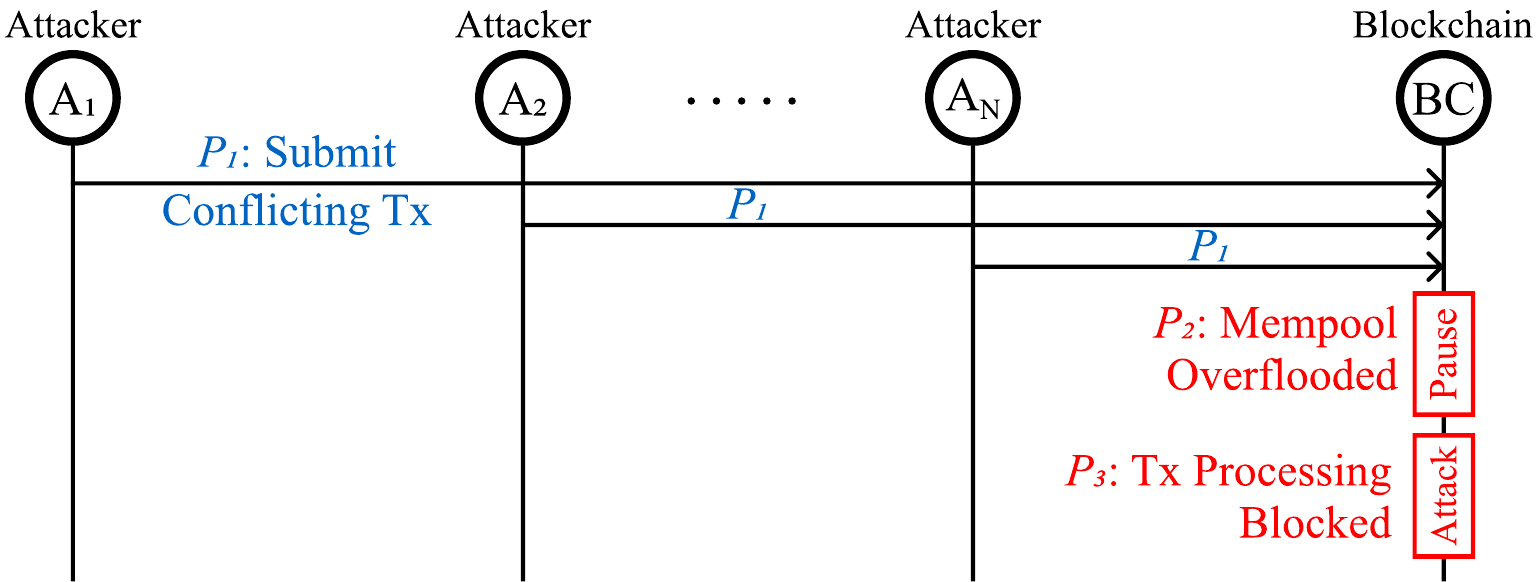}
        \caption{Sequence Diagram of DDoS Attack} \label{fig:ddos-pnet}
\end{figure}

\textbf{Permissionless Model.}
The attack's success relies on creating artificial conflicting transactions. For permissionless blockchains, careful planning is essential since each transaction incurs costs. The attacker may observe the network to generate complex conflicting transactions involving multiple wallets or launch a series of attacks to easily paralyze the network without incurring substantial costs. 

\begin{figure*}[tbh]
\centering
\includegraphics[width=0.96\textwidth]{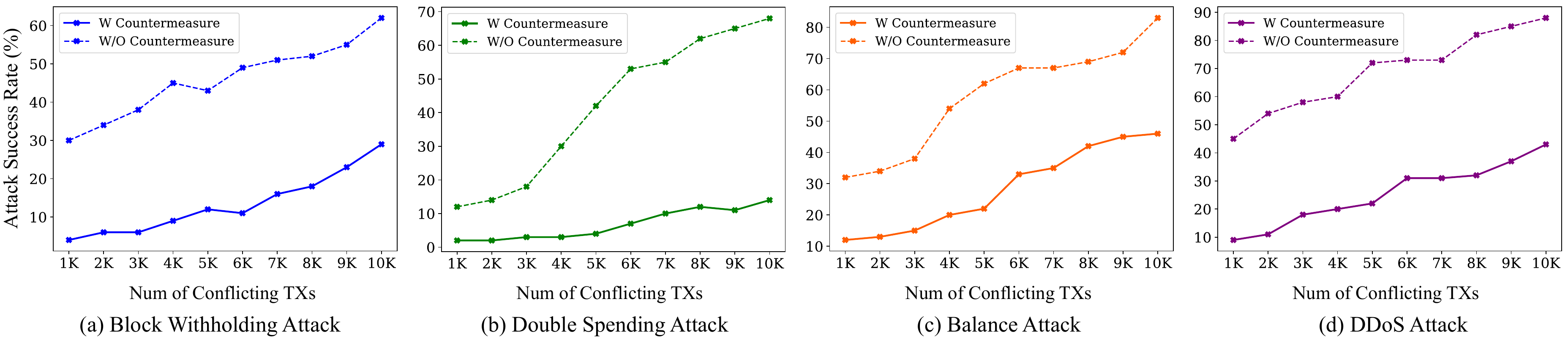}

\caption{Comparison of Attack Success rates before and after countermeasures.  } \label{fig:evaluation}
\end{figure*}

\textbf{Permissioned Model.}
In permissioned blockchains, the attacker must obtain endorsements for each transaction. This necessitates generating transactions that closely mimic real ones in the network to ensure endorsements are obtained without suspicion.

\textbf{Workflow.} 
The attacker uses numerous user accounts to submit a substantial volume of synthetically generated conflicting transactions, surpassing the typical mempool or ordering queue size. This causes the entire transaction process to stall, leading to timeouts or incorrect order execution for many valid transactions. Figure \ref{fig:ddos-pnet} shows the sequence diagram of the attack, where $P_1$ and $P_2$ are the pre-conditions for the attack while $P_3$ is the post-condition for the attack. $\bm{P_1}$: In this step, the attacker monitors pending transactions and generates conflicting synthetic transactions. These are then simultaneously submitted from 32 user accounts.
$\bm{P_2}$: After submitting those conflicting transactions, a large number of them will be waiting on the mempool and exceed the default size. If the mempool size is exceeded by those transactions, the attack is successful. $\bm{P_3}$: Now as a result of the attack, the transaction processing is blocked because of extreme latency. Also, a lot of valid transactions fail due to the inconsistent ordering.

\textbf{Outcome.}
After the successful DDoS attack, all nodes in the Fabric network experience severe latency and transaction failures. Depending on the environment and number of nodes, it may take some time to restore the network to a stable state. However, the transactions impacted by the attack will need to be resubmitted.

\section{Countermeasures}
\label{counter}
All four attacks that we modeled in Section \ref{attack-model} share a common strategy: exploiting conflicting transactions. Rather than addressing each attack individually, our focus was on tackling the root cause. In this section, we explore four countermeasures applicable to any blockchain network to mitigate these attacks.

\textbf{[C1] Checking dependency before ordering.}
To address the core issue of conflicting transactions, we implemented a validation step that checks for dependencies before ordering a transaction. This approach allows the system to identify and prevent transactions that may conflict with currently executed ones. While locking transactions and wallets has been explored as a method to avoid conflicts in ordering \cite{xu2019locking}, it can also increase network latency and disrupt regular tasks.

\textbf{[C2] Implementing priority-based ordering.}
Assigning priority levels to transactions allows the system to establish their processing order. By prioritizing transactions according to factors like type, importance, or transaction size, the network can optimize processing to minimize conflicts and ensure smoother execution. In read-heavy systems, prioritizing read operations effectively reduces performance bottlenecks \cite{nasirifard2019fabriccrdt}. We implemented a priority assigner that sets the priority based on transaction types (Read/Write). 

\textbf{[C3] Leveraging parallel processing for enhanced throughput.}
One problem that arises from ordering transactions based on priority and dependency is low throughput. To resolve this, we used parallel processing in transaction ordering. Parallel processing allows the blockchain network to execute multiple transactions simultaneously \cite{amiri2019parblockchain, gorenflo2020fastfabric} by distributing the workload across multiple processing units or nodes.

\textbf{[C4] Managing an individual transaction queue for every orderer.}
While using parallel processing, there is still a chance of conflicts between two parallel orderers. To resolve that, we used a queue for each orderer. Implementing a queue mechanism enables the orderly management of incoming transactions. It ensures that transactions are processed in a sequential and controlled manner, reducing the likelihood of conflicts and providing a structured framework for transaction execution.



\section{Implementation and Evaluation}
\label{evaluation}

To measure the efficacy of the solutions, we have implemented all four countermeasures combined in a Fabric network \cite{HyperledgerFabric}. We changed the ordering paradigm by implementing a dependency checker which only processes a transaction if all dependencies are available. We also set priority for read operations. The reason for this is most blockchain networks are read-heavy \cite{chacko2021my}. Moreover, we implemented parallel processing in the ordering phase with individual orderers maintaining a queue. 

\begin{figure}[tbh]
\centering
\includegraphics[width=0.9\columnwidth]{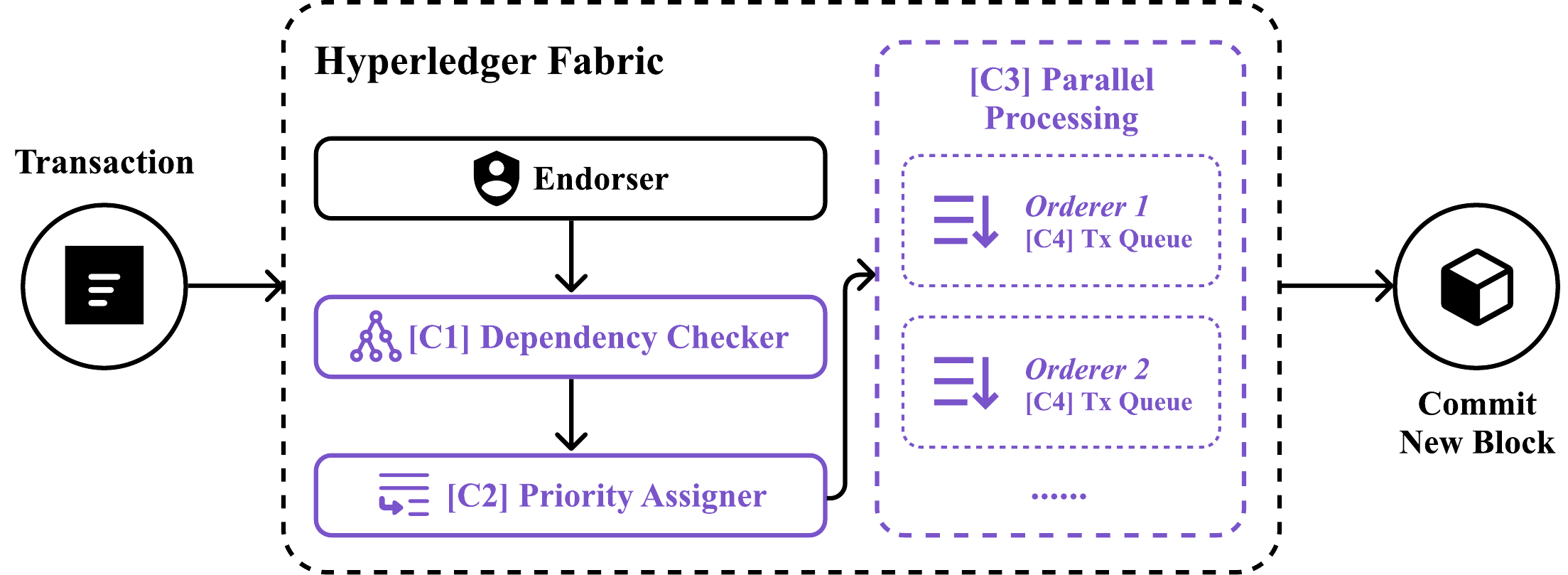}
        \caption{Modified Architecture of Hyperledger Fabric after adding Countermeasures} \label{fig:arc}
\end{figure}

The architecture of the modified Hyperledger Fabric with countermeasures is presented in Figure \ref{fig:arc}. After implementing all these countermeasures for handling conflicting transactions, we simulated all four attacks 100 times for different numbers of conflicting transactions ranging from 1000 to 10000. Figure \ref{fig:evaluation} shows the effectiveness of the countermeasures for these attacks. The dashed lines represent the success rate in a default Hyperledger Fabric network. For most of the attacks, the usual success rate was relatively higher with a maximum of 90\% for 10000 conflicting transactions. However, after applying the countermeasures, we were able to reduce the attack rate significantly with a maximum of 42\% for 10000 conflicting transactions. And for the low number of conflicting transactions, the success rate remains lower than 10\% for all of the attacks. 

\begin{table}[]
\centering
\caption{Overhead and Throughput Comparison}
\label{tab:overhead-result}
\begin{tabular}{lcc}
\hline
                    & \textbf{No Countermeasures} & \textbf{Countermeasures} \\ \hline
Avg. Memory Usage   & 4.73 GB                  & 4.95 GB                 \\ \hline
Max Memory Usage    & 6.95 GB                  & 7.31 GB                 \\ \hline
\textcolor{red}{Memory Overhead}     & \multicolumn{2}{c}{\textcolor{red}{$\sim$5\% Increased}}                      \\ \hline
Avg. Throughput     & 150 TPS                  & 1800 TPS                \\ \hline
\textcolor{blue}{Throughput Increase} & \multicolumn{2}{c}{\textcolor{blue}{12X Faster}}                   \\ \hline
\end{tabular}
\end{table}

Moreover, we also measured the memory overhead after implementing the countermeasures. For the experimental setup, we used four virtual machines running a blockchain node with 8 GB of memory each. Table \ref{tab:overhead-result} shows the comparison of memory usage and throughput before and after adding the countermeasures. The implementation of such countermeasures adds nearly 5\% memory overhead. However, despite the extra memory usage, the network's performance undergoes a uplift as we deploy parallel processing. While initially averaging 150 transactions per second (TPS), throughput surges to 1800 TPS post-parallel processing.

\section{Related Works}
\label{related}
Several research studies explore attacks and defenses within blockchain technology. Saad et al. \cite{saad2020exploring} examine the attack surface of blockchain, noting vulnerabilities like selfish mining, 51\% attacks, and smart contract exploits while suggesting defense measures. However, they acknowledge that attacks can occur through different methods. Zhang et al. \cite{zhang2019double} introduce a new attack model combining double-spending with Sybil attacks in the Bitcoin network, focusing on block propagation disruption. Similarly, Ramezan et al. \cite{ramezan2018strong} propose an adaptive strategy for double-spending attacks, showing higher success probabilities. Tosh et al. \cite{tosh2017security} and Lee et al. \cite{lee2019countering} showed block withholding attacks in cloud and PoW-based cryptocurrencies. Natoli et al. \cite{natoli2017balance} identify the balance attack against forkable blockchain systems, advocating for non-forkable designs to mitigate the threat. While these studies improve blockchain security by identifying vulnerabilities and proposing effective defense mechanisms across various application domains, they do not address vulnerabilities stemming from conflicting transactions.

On the other hand, some researchers have explored on the impact of conflicting transactions on the performance of blockchains. They mostly focused on the transactions per second and latency which led to some high-performance frameworks like XOX Fabric \cite{gorenflo2020xox}, Fabric CRDT \cite{nasirifard2019fabriccrdt}, and ParBlockchain \cite{amiri2019parblockchain}. However, none of these schemes actually explored the security vulnerabilities that may arise from conflicting transactions. We are the first to explore the impact of conflicting transactions in blockchain attack vectors offering innovative mitigation techniques to enhance security and performance concurrently.

\section{Conclusion}
\label{conclude}
In this paper, we have provided a comprehensive exploration of the impact of conflicting transactions on blockchain attack vectors. Using attack models, we demonstrated how attackers can meticulously generate a large number of conflicting transactions to cause transaction ordering inconsistencies and launch major attacks. Although these attacks can be executed in numerous ways and have different mitigation strategies, we are the first to explore the impact of conflicting transactions and propose countermeasures for them. The evaluation results demonstrate that effective handling of conflicting transactions significantly reduces the possibility of attacks with minor memory overhead. Furthermore, mitigating conflicting transactions yields notable performance enhancements, emphasizing the relevance of our presented countermeasures in this paper.

\bibliographystyle{IEEEtran}
\bibliography{IEEEabrv,references}

\end{document}